\documentclass[a4paper,11pt]{article}
\pdfoutput=1
\usepackage{jinstpub}
\usepackage{natbib}
\usepackage{appendix}
\usepackage{xcolor}
\newcommand{\code}{\texttt}
\begin{document}

%% TITLE
%\title{Ray tracing the FOXSI rocket optics and application for ghost ray analysis}
%\title{Validation of a ray tracing simulation for the FOXSI rocket experiment}
\title{Use of a ray-tracing simulation to characterize ghost rays in the FOXSI rocket experiment}
%\title{FOXSISIM ray tracing simulation. Study case: ghost rays for the FOXSI rocket payload}
%\title{Characterization of the FOXSI rocket optics: A study case for the FOXSISIM ray tracing simulation}

%% AUTHORS
\author[a]{J.C. Buitrago-Casas,} % ORCID ID 0000-0002-8203-4794
\author[b]{S. Christe,} % ORCID ID 0000-0001-6127-795X
\author[c]{L. Glesener,}
\author[a,d]{S. Krucker}
\author[e]{B. Ramsey,}
\author[e]{S. Bongiorno,}
\author[e]{K. Kilaru,}
\author[c,e]{P.S.Athiray,}
\author[f]{N. Narukage,}
\author[g]{S. Ishikawa,}
\author[a]{G. Dalton,}
\author[a]{and S.Courtade}
\author[c,h]{S. Musset,}
\author[c]{J. Vievering,}
\author[b]{D. Ryan,}
\author[a]{and S. Bale}

%% AFFILIATIONS
\affiliation[a]{University of California Berkeley, Space Sciences Laboratory, 7 Gauss way, Berkeley, USA, CA 94720}
\affiliation[b]{NASA Goddard Space Flight Center, 8800 Greenbelt Rd, Greenbelt, USA, MD 20771}
\affiliation[c]{University of Minnesota, Physics \& Astronomy, 116 Church St. SE, Minneapolis, USA, MN 55455}
\affiliation[d]{University of Applied Sciences Northwestern Switzerland, Windisch, Switzerland}
\affiliation[e]{NASA Marshall Space Flight Center, Martin Rd SW, Huntsville, USA, AL 35808}
\affiliation[f]{National Astronomical Observatory of Japan, 2 Chome-21-1 Osawa, Mitaka, Tokyo 181-0015, Japan}
\affiliation[g]{Rikkyo University, Graduate School of Artificial Intelligence and Science, 3-34-1 Nishi-Ikebukuro, Toshima, Tokyo 171-8501, Japan}
\affiliation[h]{SUPA, School of Physics \& Astronomy, University of Glasgow, Glasgow G12 8QQ, UK}
%% EMAIL
\emailAdd{milo@ssl.berkeley.edu}

%% ABSTRACT
%\input{sections/00_Abstract.tex}

\abstract{
Imaging X-rays by direct focusing offers greater sensitivity and a higher dynamic range compared to techniques based on indirect imaging. The Focusing Optics X-ray Solar Imager (FOXSI) is a sounding rocket payload that uses seven sets of nested Wolter-I figured mirrors to observe the Sun in hard X-rays through direct focusing. Characterizing the performance of these optics is critical to optimize their performance and to understand their resulting data. In this paper, we present a ray-tracing simulation we created and developed to study Wolter-I X-ray mirrors. We validated the accuracy of the ray-tracing simulation by modeling the FOXSI rocket optics. We found satisfactory agreements between the simulation predictions and laboratory data measured on the optics. We used the ray-tracing simulation to characterize a background pattern of singly reflected rays (i.e., ghost rays) generated by photons at certain incident angles reflecting on only one of a two-segment Wolter-I figure and still reaching the focal plane. We used the results of the ray-tracing simulation to understand, and to formulate a set of strategies that can be used to mitigate, the impact of ghost rays on the FOXSI optical modules.  These strategies include the optimization of aperture plates placed at the entrance and exit of the smallest Wolter-I mirror used in FOXSI, a honeycomb type collimator, and a wedge absorber placed at the telescope aperture. The ray-tracing simulation proved to be a reliable set of tools to study Wolter-I X-ray optics. It can be used in many applications, including astrophysics, material sciences, and medical imaging.
}


%% ArXiv number:
\arxivnumber{1234.56789} %% Need to edit this when submitted
\collaboration{On behalf of the FOXSI collaboration.}
\maketitle
\flushbottom

%% CONTENT
\section{Introduction}
\label{sec:intro}

\begin{figure}[htbp]
\centering % \begin{center}/\end{center} takes some additional vertical space
\includegraphics[width=1.0\textwidth]{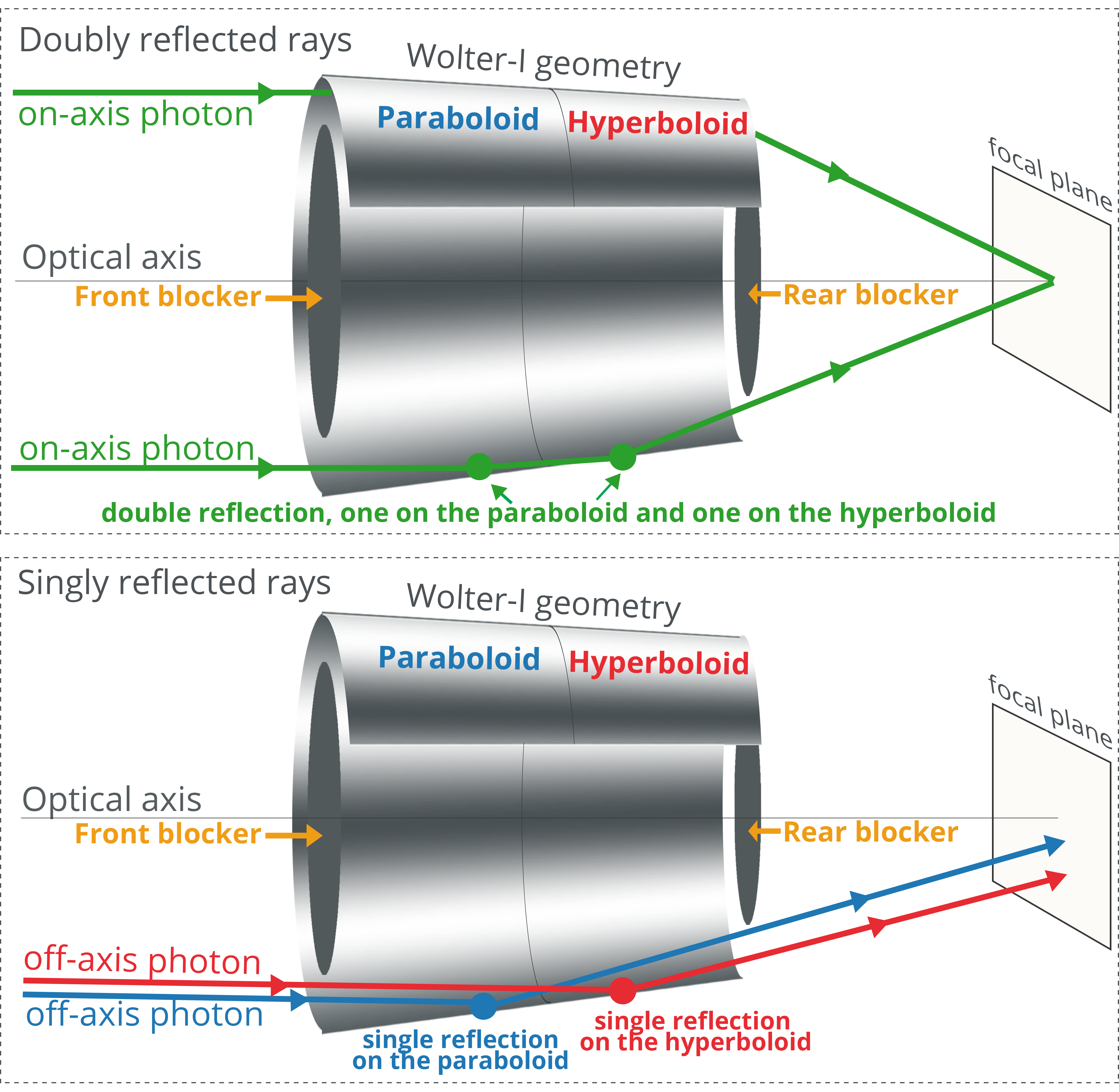}
\caption{\label{fig:mirrors} A Schematic of a Wolter-I monolithic mirror shell with the parabolic and hyperbolic reflecting surfaces showing the difference between on-axis rays (green) that reflect on both mirror surfaces and those off-axis rays (red and blue) that only reflect off of a single mirror surface. The optical axis is depicted as a perpendicular line to the focal plane that goes through the center of the optics. In the {\bf top} panel, on-axis photons reflect first on the paraboloid segment then on the hyperboloid section and come to a focus on the focal plane. These are referred to as doubly reflected focused rays. Blockers are primarily used (front and rear indicated by the yellow arrows) to block rays which would go straight through the module without reflecting off either surface and reach the focal plane. In the {\bf bottom} panel,  Singly reflected rays coming from off-axis angles interact only with a single mirror surface (either the paraboloid, blue rays, or hyperboloid, red rays, segment) and can make it to the focal plane. These singly-reflected rays are frequently referred to as ghost rays because they can lead to unfocused patterns on the focal plane. The blockers can also reduce the amount of ghost rays by limiting the angles accessible by off-axis rays to reflect off of mirror segments.}
\end{figure}

Since first introduced in 1950 \cite{wolter1950frage}, Wolter-I-figured grazing incidence X-ray telescopes have been extensively used in a variety of areas such as synchrotron accelerators, nuclear physics, astrophysics, and space physics (e.g. \cite{cash2002medical,ogasaka2008characterization,mildner2011wolter,ferreira2013hard}).  The Wolter-I geometry consists of a combination of two grazing-incidence mirror segments, a paraboloid primary mirror followed by a hyperboloid secondary reflector, referred to as a mirror shell. Frequently, to build up effective area, many individual mirror shells with different diameters are nested together to form a telescope module. On-axis rays which reflect off both mirrors are brought to a focus to form an image on the focal plane. There exists the possibility though, that rays from off-axis sources may reflect only on a single surface and reach the focal plane. These single-reflecting rays are frequently referred to as stray light, or ghost rays (see Figure \ref{fig:mirrors}) \cite{werner1977imaging}. 

%% Ghost Rays - Singly and doubly reflected rays

%% Ghost Rays - Singly and doubly reflected rays
\begin{figure}[htbp]
\centering % \begin{center}/\end{center} takes some additional vertical space
\includegraphics[width=1.0\textwidth]{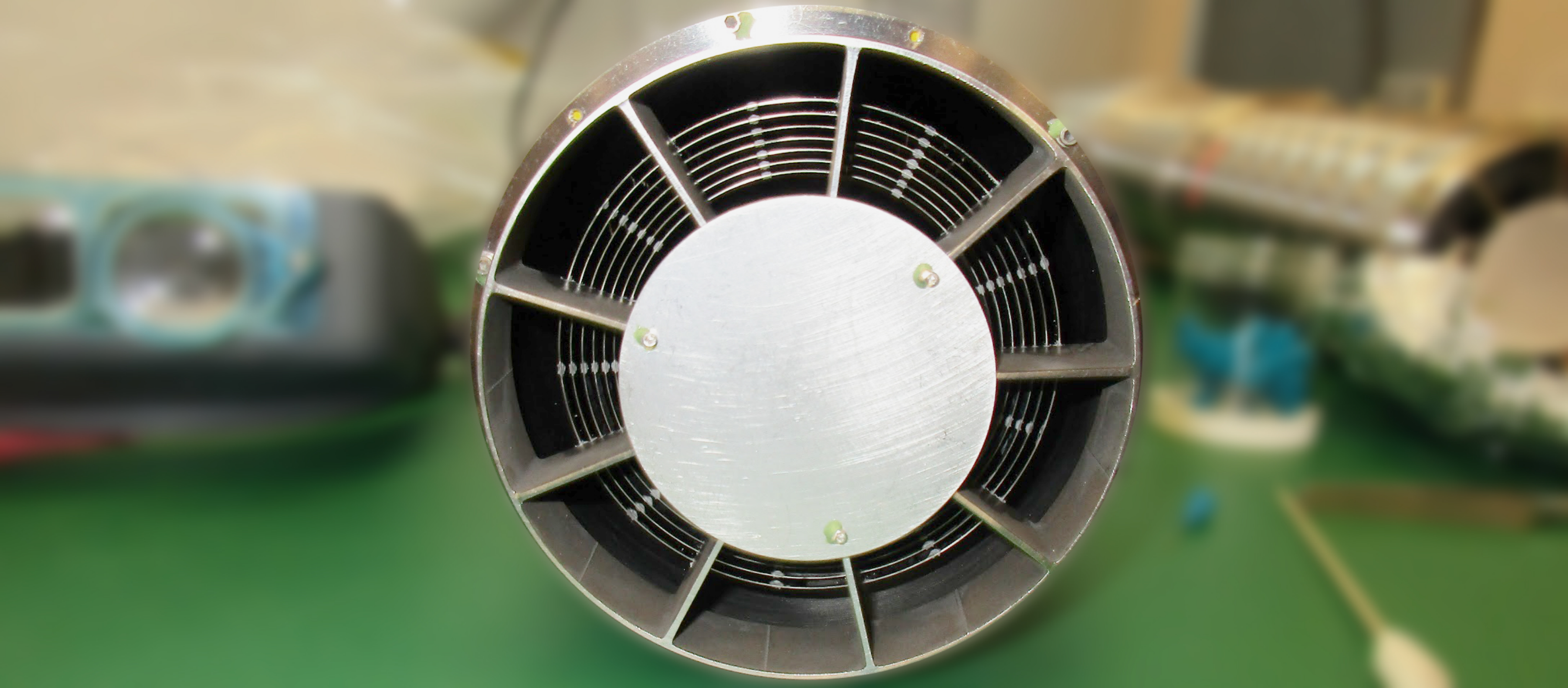}
\caption{\label{fig:photograph} Photograph of the front of a FOXSI sounding rocket telescope module nesting 7 Wolter-I mirrors shells. The circular front blocker is clearly visible and a spoked structure is used to hold and maintain alignment of each mirror shell with respect to the others.
}
\end{figure}

Ghost rays are generally not an issue when observing isolated on-axis point sources since there are then no off-axis sources. This is frequently the case for astrophysical sources; most X-ray sources are far enough away from us that they are point sources and are far enough apart (many arcmin) from each other that they can be considered isolated. However, when observing the Sun, whose angular diameter is $\sim$0.5 deg, it is possible to have multiple bright X-ray sources simultaneously emitting within a relatively small angular extent on the sky. Some of these sources can be off-axis when pointing at a target of interest and can generate ghost rays which may obscure the emission from the primary target, deteriorating the performance of a solar telescope. Previous missions have minimized ghost rays by, for instance, tightly packing and nesting the mirror shells to block the paths of ghost rays as in the case of the Nuclear Spectroscopic Telescope Array ({\it NuSTAR}) \cite{madsen2017observational}, through a set of concentric circular sieves at the telescope aperture, in the case of Simbol-X \cite{cusumano2007simbol}, or by using a pre-collimator as described by \citet{spiga2016analytical}. In this paper, we describe a number of approaches to mitigate ghost rays that were developed, simulated, and tested for the FOXSI sounding rocket payload. Since the FOXSI science goals require  high sensitivity and dynamic range, understanding and mitigating ghost rays is crucial to substantially improve its science. In order to verify these approaches a new open source ray tracing simulation tool was developed which is also introduced here.

\section{The \code{foxsisim} ray-tracing Simulation}
\label{sec:simulation}

\begin{figure}[htbp]
\centering 
\includegraphics[width=1.0\textwidth]{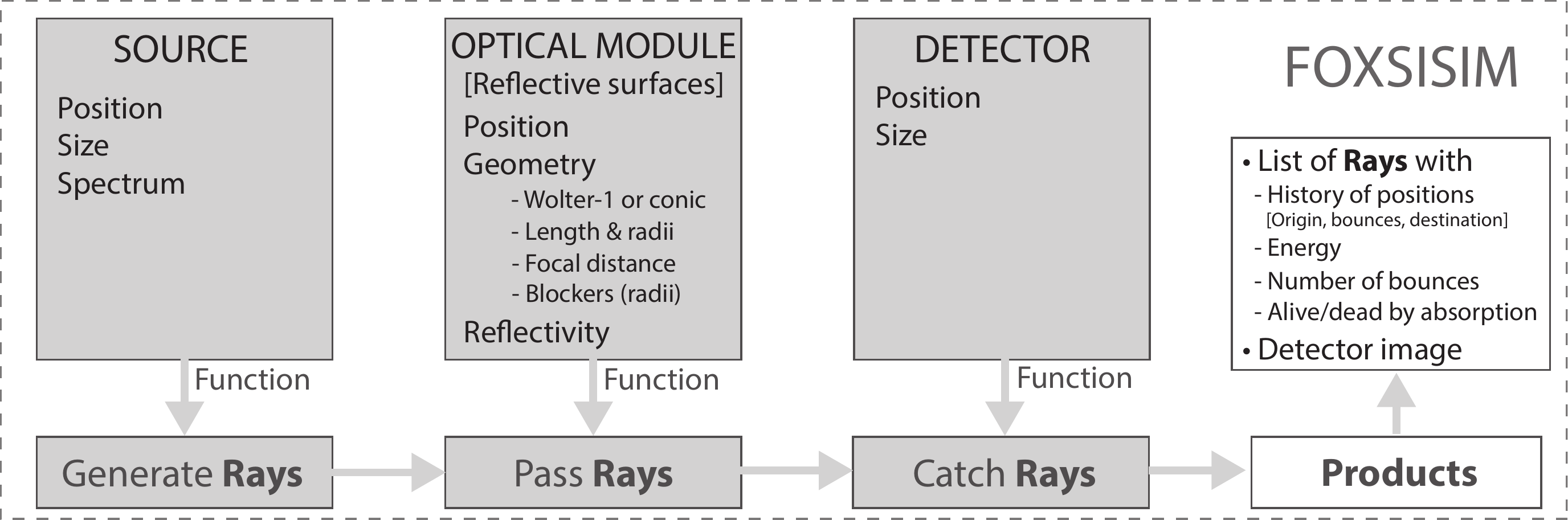}
\caption{\label{fig:FlowDiagram} The flow diagram for the functional structure of the \code{foxsisim} ray-tracing simulation tool. The code has three basic classes: \code{Source}, the source of x-rays, optical \code{Module} (a telescope module), and a \code{Detector}. For each component in the simulation, a set of initialization parameters need to be defined. Some examples are shown here which include position, size, spectrum of a source or reflectivity of an optical surface. \code{Source} implements a function to generate a list of random {\bf rays}. The \code{pass rays} function defined by \code{Module} computes the interactions of rays with the module. Finally, the  \code{Detector} class implements a \code{catch} function which computes which rays land on the detector. The final output is a list of all rays with keywords defining their position histories, their energy, their number of reflections, and a tag indicating whether or not a ray was absorbed on a non-reflecting surface.
}
\end{figure}

In order to verify and validate the properties of grazing-incidence optics and ghost-ray mitigation strategies, we created and optimized a ray-tracing simulation capability. This tool is referred to as \code{foxsisim}, though it is generic and can be used for many different grazing-incidence configurations. Written in Python, the source code is open source and available on Github\footnote{\url{https://github.com/foxsi/foxsi-optics-sim}}\citep{steven_d_christe_2019_3445460}. The software is provided as a Python package with documentation and also provides a simple graphical user interface. The \code{foxsisim} package provides a class-based approach to ray-tracing. Figure \ref{fig:FlowDiagram} shows a flow diagram representing the functional structure of \code{foxsisim}.

%% Flow Diagram

The \code{Module} class uses a \code{Surface} subclass, which represents any optical surface with which rays may interact. Such a surface is generally represented by a parametric equation in 3 dimensional space. The \code{SegmentH} and \code{SegmentP} are subclasses of \code{Surface} and represent hyperboloid and parabolic mirror segments respectively. A \code{Shell} class holds both reflecting segments and therefore represents a telescope shell. Finally a full telescope module is potentially composed of many shells and is represented by a \code{Module} class. An X-ray source is implemented in the \code{Source} class which implements both isotropic sources as well as sources of perfectly parallel beams. The geometric rays which represent the X-rays themselves are implemented by the \code{Ray} class. A casting function finds the solution of rays reflecting off of or being stopped by surfaces (reflecting or non-reflective). The spectral properties of the telescope are computed through its reflectivity, which depends on the incoming ray angle. \code{foxsisim} uses Iridium surfaces with 10\AA$\,$ roughness to compute reflectivities. Finally, a  \code{Detector} class is used to capture the rays at a particular point in space and create an image. The current implementation does not consider any random scattering off of surfaces, detector efficiency, nor prioritize speed, though a future release is planned to address these issues.

\section{The FOXSI sounding rocket payload and ghost rays}
\label{sec:FOXSI-Ghost-rays}

The FOXSI sounding rocket program is a mission to develop and test grazing-incidence optics for solar observations. 
FOXSI uses a set of 7 Wolter-I figured grazing-incidence X-ray telescope modules to perform imaging spectroscopy of solar hard X-rays from $\sim 4$keV to $\sim20$keV \cite{krucker2009focusing,krucker2013focusing}.  The parameters of the optics, such as diameters and focal length, were set to suit the payload of a Terrier-Black-Brant sounding rocket. These optics were produced at NASA Marshall applying a low-cost electroformed nickel alloy replication process, whereby nickel mirrors are electro-deposited onto super-polished mandrels. For increased effective area, shells of various radii are coaxially nested together into modules of 7 or 10 mirrors. The averaged resolution of the integrated modules was measured in the laboratory to be 4.3±0.6 arcsec (FWHM) and 27±1.7 arcsec (half-power diameter; HPD) for an on-axis source (\cite{christe2016JAI, krucker2013focusing}. A primary advantage of this production method NASA Marshall uses is that multiple mirrors can be fabricated via replication by using a single mandrel, which substantially lowers the cost of producing identical optics.
A detailed description of how these optics are fabricated can be found in \cite{Ramsey2002} and \cite{Ramsey2005}. The payload uses seven 60 cm long telescope modules, each one containing 7 or 10 concentrically nested Wolter-I mirror shells, all with a focal length of 2 m. The mirrors are optimal for graze angles ranging from 0.23 to 0.37 degrees. Constrained to the FOXSI Si detector square area, the field of view (FOV) is 16 $\times$ 16 arcmin$^2$. Each optical module includes circular blockers at the front and rear apertures added as a baseline to mitigate straight-through and some ghost rays (see diagram and photograph in Figures \ref{fig:mirrors} and \ref{fig:photograph}, respectively). The circular blockers are 1.55 mm thick aluminum disks with 37.5 mm (front) and 31.4 mm (rear) radii for a 7-mirror module, and with 31.0 mm (front) and 26.2 mm (rear) radii for a 10-mirror module.

A number of papers in the literature provide background on this experiment. \citet{krucker2013focusing,krucker2014first} describe the original payload and first scientific results of the mission. \citet{Glesener:2016eq} provides an overview of the first two flights of the experiment. \citet{christe2016JAI} describes major updates made for the second flight as well as details on the mirror shell prescription. \citet{Musset:2019gj}, \citet{athiray2017calibration} and \citet{furukawa2019development} describe the hardware upgrades for the 3rd flight of the sounding rockets. This paper is a continuation of the work described in \citet{BuitragoCasas:2017dy} which describes implementations  to reduce singly-reflected X-rays. Outstanding sciences results based on FOXSI's observations are reported by \citet{ishikawa2017detection,athiray2020foxsi}.

FOXSI's science objectives require high sensitivity and dynamic range observations of the Sun; therefore understanding and mitigating ghost rays is essential. In order to characterize the patterns generated by off-axis sources, a number of measurements were performed on individual telescope modules at the NASA Marshall Stray Light Facility (SLF). The SLF consists of a 100 m evacuated beam line. An optics module is placed inside an evacuated chamber at one end of the facility. Two stages enable the optic to be rotated with a precision down to $\sim3.5$ arcsecs. The optics module is then illuminated by a Trufocus 50 keV X-ray generator with a molybdenum target located at the other end of the facility. A cooled Andor CCD iKon-L camera is located at the focal plane. The Andor CCD camera consists of a 204y8$\times$2048 array of pixels, with a pixel pitch of 13.5 microns. This pitch translates to a resolution of 1.3 arcseconds and a measurement field of view (FOV) of 44.4$\times$44.4 arcminutes, much larger than the FOXSI detector field of view. Since the X-ray source cannot be moved, the optic is rotated to simulate rays coming from off-axis positions.

%% Ghost Rays Color Coding
\begin{figure}[htbp]
\centering
\includegraphics[width=1.0\textwidth]{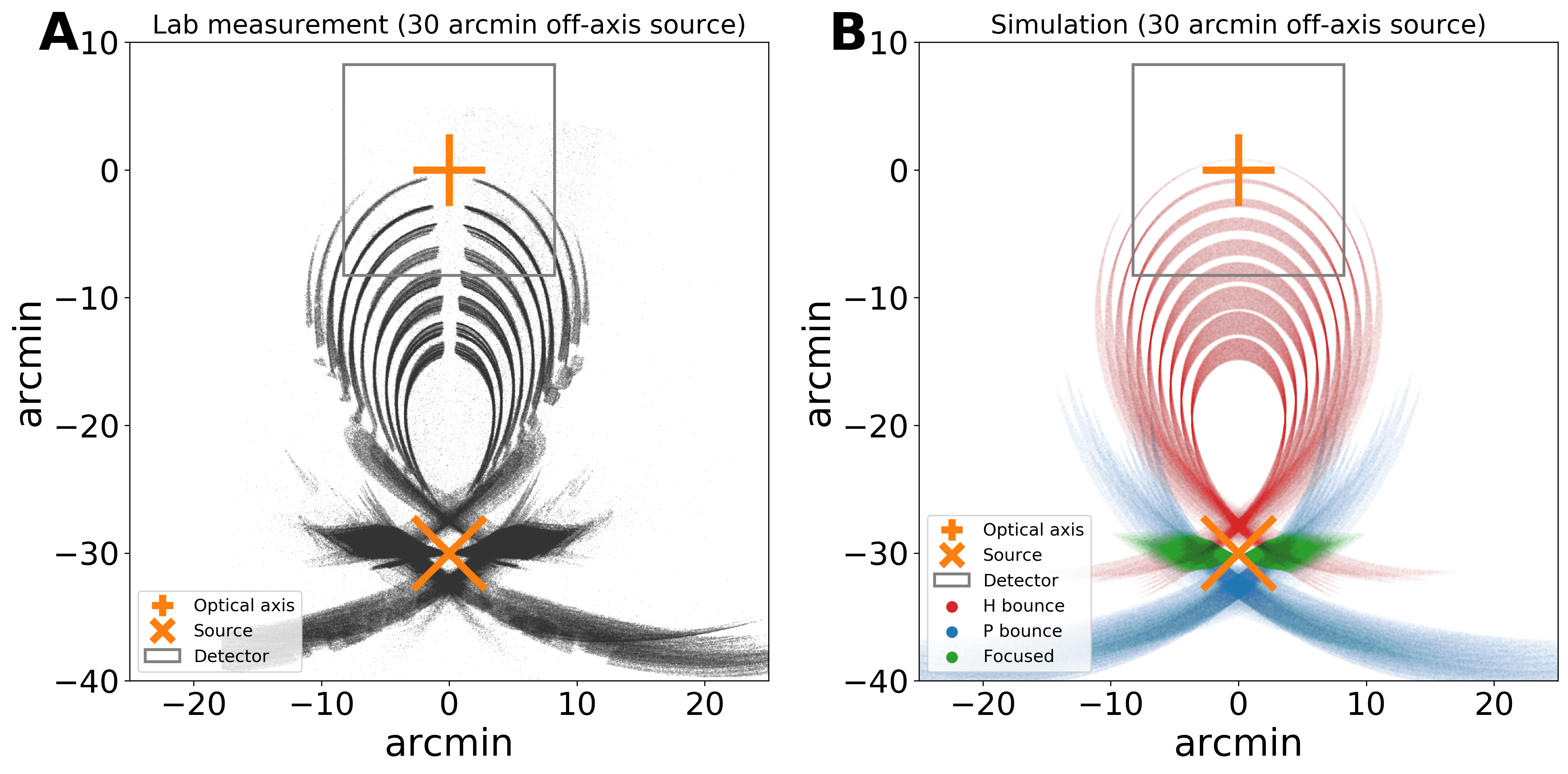}
\caption{\label{fig:sim-data}
The ghost ray image from an X-ray source 30 arcmin off-axis for a 10-shell optics module. In both panels, the grey square represents the on-axis FOXSI field of view. At the center, the optical axis is shown as the orange cross. The location of the off-axis source is denoted by the orange X. Panel {\bf A}: A measured ghost ray image produced by a 10-shell optics module for a source located 30 arcmin off-axis. Data were acquired at the NASA MSFC 104 m Stray Light Facility using a 2048 by 2048 pixel CCD detector placed at the focal plane. Panel {\bf B}: Simulated ghost rays produced by the \code{foxsisim} ray-tracing code. The green portion of the pattern corresponds to doubly-reflected rays while the blue and red areas are the patterns generated by the paraboloid and the hyperboloid segments, respectively.
}
\end{figure}

Figure \ref{fig:sim-data} shows an example of a ghost ray pattern from an off-axis X-ray source 30 arcmin away from the optical axis. Panel A shows an actual measurement taken at the SLF with the Andor CCD camera. The gray box shows the limited FOXSI field of view with the optical axis (orange cross) at its center. The position of the off-axis source is shown by an orange {\it X} mark. A complex ghost ray pattern was measured which is a result of singly-reflected rays from both the hyperboloid and paraboloid mirror shell segments. In this case this is a 10-shell telescope. Each shell creates an individual ring pattern which extends onto the FOXSI field of view. Notches in the rings are caused by the spider spokes which hold the shells seen in Figure \ref{fig:photograph}.

Panel B in Figure \ref{fig:sim-data} shows the results of a \code{foxsisim} ray-tracing simulation which compare well with the measured ghost ray pattern. The simulated image is color-coded to show rays that are doubly-reflected (green), reflected only by the hyperbolic segments (red), and reflected only by the parabolic segments (blue). The major features in the ghost ray patterns are rings, one for each of the ten mirror shells. The vertical gap in the rings are caused by one of the spider spokes (seen in Figure \ref{fig:photograph}) which partially block the shells. These are not included in the simulation. The blockers are included in the simulation and block some of the ghost rays. A first order quantifiable comparison of the lab data with the simulation was done by measuring the distance from the source location to each of the rings in the patterns. Such comparison shows differences always under $\sim 3\%$ for all the rings of the lab data compared to the ones for the simulation.

\section{Use of \code{foxsisim} to study FOXSI ghost-rays}
\label{sec:foxsisim-ghostrays}

\subsection{Ghost rays as function of off-axis angle}
\label{subsec:emergence}

\begin{figure}[htbp]
\centering
\includegraphics[width=1.0\textwidth]{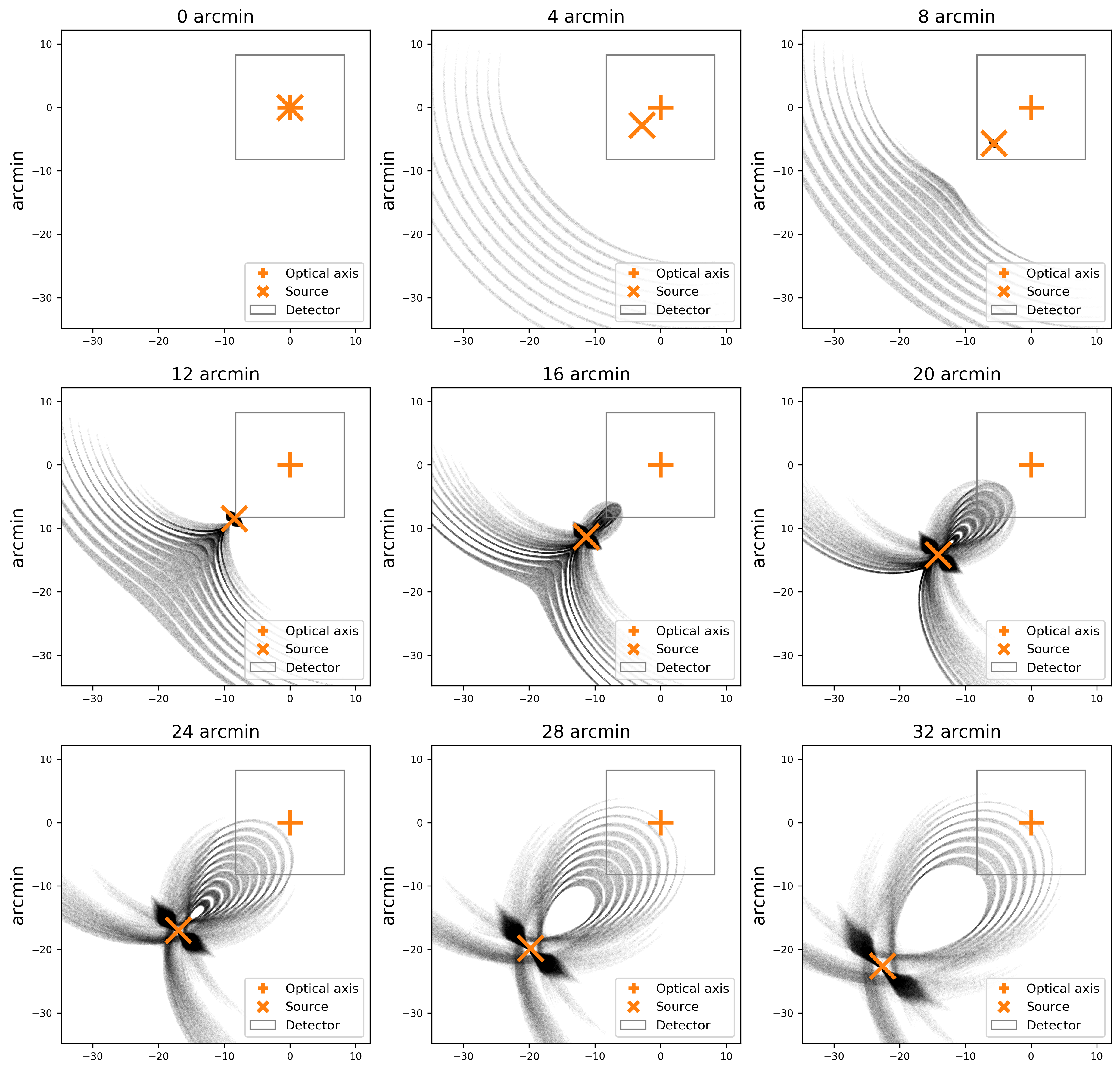}
\caption{\label{fig:GRemergence} 
Simulated ghost ray images for a 10-shell module as a function of off-axis angle for a source at infinity from 0 to 32 arcmin. The gray square shows the field of view of the FOXSI detectors. The orange symbols show the optical axis and source position. As the source moves away from the optical axis ghost ray patterns appear, at first outside of the detector-bounded field of view. Between 12 and 16 arcmin, these rays begin to infringe on the detector area. The detector-integrated ghost ray flux contamination continues to increase as the pattern increases in size and complexity.}
\end{figure}

With the simulation validated and in order to gain a better intuition for how ghost rays evolve as a function of off-axis angle, the simulation was run for a variety of off-axis angles. The results are shown in Figure~\ref{fig:GRemergence} for off-axis angles 0, 4, 8, 12, 16, 20, 24, 28, and 32 arcmin away from the optical axis. It can be seen that as soon as the source is moved off-axis, a ghost-ray pattern emerges but at large angular distances away from the focal plane. These rays do not (yet) infringe on the detectors plane. As the source moves further off-axis, these ghost rays change shape and eventually begin to be seen on the focal plane between 12 and 16 arcmin. A more precise analysis shows that the threshold for ghost rays to impact the detect plane is at 13 arcmin off-axis for a 10-mirror optics module, and 18 arcmin off-axis for a 7-mirror optics module. This difference is due to the fact that most ghost rays originate  from the inner most mirrors of the optics. A 10-mirror optics module includes three additional smaller mirrors to the standard configuration of a standard 7-mirror optics module, as described in \citet{christe2016JAI}.

%% Ghost Rays emergence with off-axis angle 

\subsection{Ghost rays energy dependence}
\label{subsec:energy-dependence}

The reflectivity of a Wolter-I mirror depends on the angle of incidence, the surface material, and roughness of the mirror \cite{attwood2017x}. The FOXSI optics uses electroformed nickel mirrors coated with a thin layer of iridium. For the simulation, we use the theoretical reflectivity of  Iridium with a surface roughness of 10\AA\footnote{\href{http://henke.lbl.gov/optical_constants/}{http://henke.lbl.gov/optical\_constants/}}. Compared to focused rays, ghost rays are generated from rays at large angles yet they are only reflected once so that it is not straightforward to understand their energy dependence. To investigate this, we simulated the energy response of ghost rays.

\begin{figure}[htbp]
\centering % \begin{center}/\end{center} takes some additional vertical space
\includegraphics[width=1.0\textwidth]{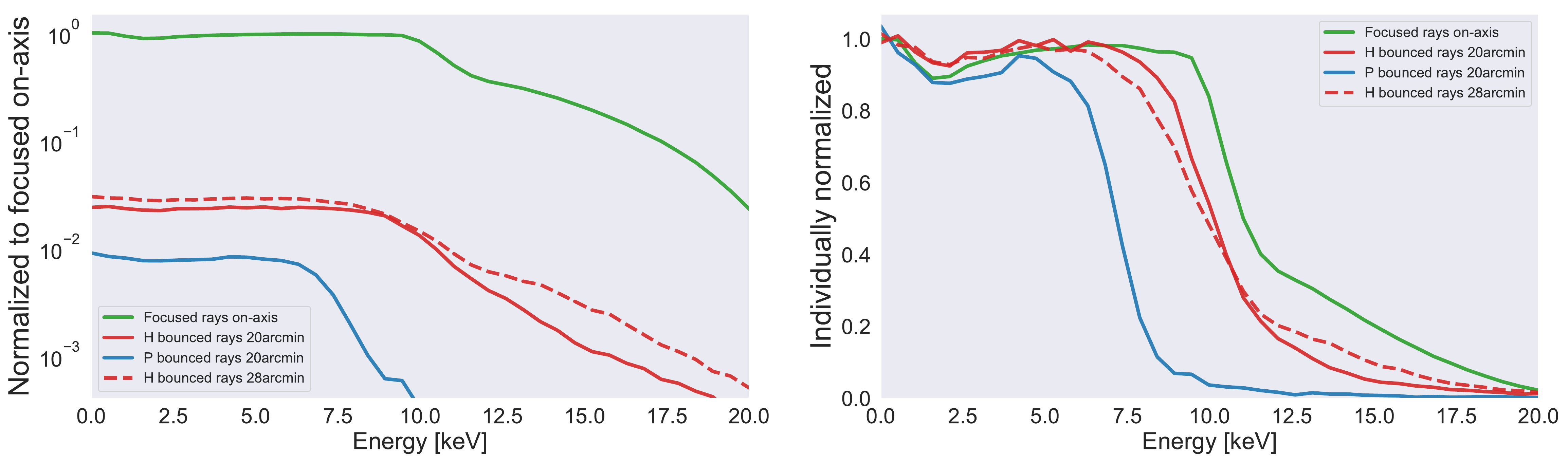}
\caption{\label{fig:GRenergy} 
The simulated energy response for a point source with a flat spectrum from 0 to 30 keV integrated over the detector field of view from an on-axis (green solid line), 20 arcmin (solid red and blue lines), and 28 arcmin (dashed red line) off-axis. Blue and red lines represent singly-reflected rays from the paraboloid and hyperboloid segment, respectively. {\it Left}. Detector-integrated fluxes normalized to the focused on-axis (solid green line) at 1 keV. The singly-reflected flux intensity is found to be two orders of magnitude smaller than the focused intensity. {\it Right}. The same curves from the left but normalized to each other to better compare the energy dependence. It can be seen that the ghost ray intensity falls off significantly faster than the focused rays. This is most pronounced for those singly-reflected rays from the paraboloid segment.}
\end{figure}

Figure \ref{fig:GRenergy} shows the simulated energy dependence of the effective area with energy for doubly- and singly- reflected rays generated by an on-axis source, an off-axis source at 20 arcmin (solid lines), and an off-axis source at 28 arcmin (dashed lines) on a 10-mirror optics integrated over the detector field of view. The input spectrum is assumed to be constant from 0 to 30 keV. On the left panel of figure \ref{fig:GRenergy}, the y-axis is normalized to focused flux at 1 keV (green curve), and it is plotted in log scale. For both panels, the solid blue (red) lines correspond to the paraboloid (hyperboloid) ghost rays reaching the detector area coming from a 20 arcmin off-axis source. The dashed red line corresponds to hyperboloid ghost rays coming from a 28 arcmin off-axis source. The left panel shows that ghost ray fluxes are generally two orders of magnitude smaller than the flux from an on-axis source. On the right, the fluxes are all normalized to each other to better compare the energy dependence. From this panel, it is clear that , above $\sim$6 keV, the ghost ray flux falls off significantly faster than for focused rays. This behavior is even more  dramatic for those rays that singly reflect off of the paraboloid. This simulation allow us to conclude that ghost rays provide only a small contribution to the background and are negligible for source-dominated observations. Furthermore their contribution to the background is reduced at high energies. 

% Rays Energy response first results. Each source was simulated to emit a total of $2.5\times10^7$ rays towards the FOXSI entrance.

An analog analysis of ghost rays dependence on off-axis angles and energy, but scaled up to a satellite mission, is presented in appendix \ref{appendix} of this paper.

\section{Ghost Ray Mitigation Strategies}
\label{subsec:strategies}

In this discussion we explore a number of different strategies to reduce the intensity of ghost rays.

\subsection{Circular Blockers}
\label{subsec:blockers}

The concept of circular blockers has already been introduced in section \ref{sec:FOXSI-Ghost-rays}. These blockers located at the entrance aperture and exit primarily stop X-rays from traveling through the center of the optics. The back aperture blocker also serves to remove some ghost rays. In this section, we investigate the optimization of the size of these blockers, both front and rear. For optimized blockers, fewer ghost rays reflected from the hyperboloid segments make it onto the detector. If not correctly sized, small blockers can allow some paraboloid reflected ghost rays to reach the detector area. On the other hand, if the blockers are too large, the focused rays from the smallest mirror are blocked. Optimizing their design is crucial to improve the overall performance of the telescope.

\begin{figure}[htbp]
\centering
\includegraphics[width=1.0\textwidth]{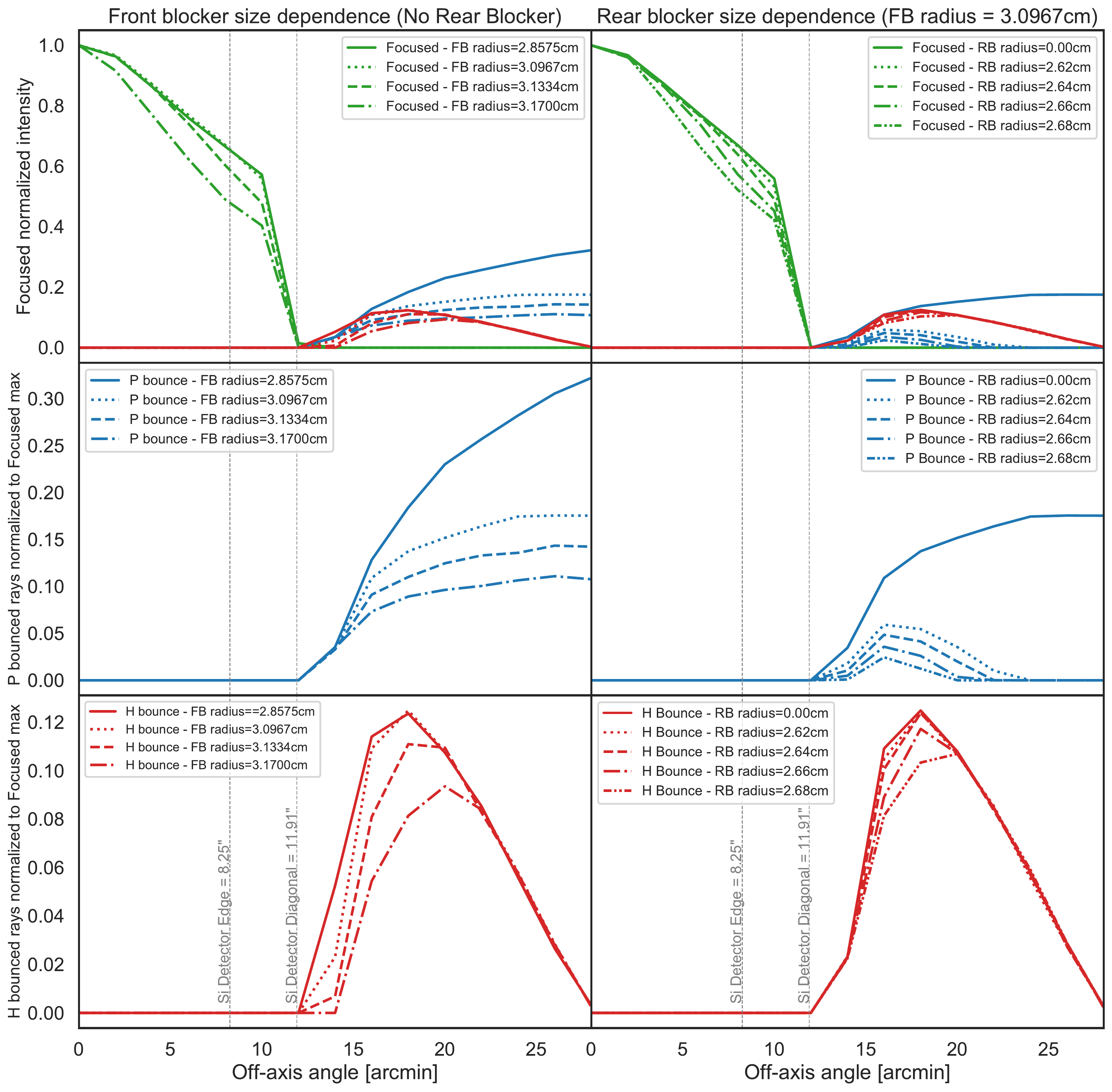}
\caption{\label{fig:blockers}
Use of \code{foxsisim} to study the effect that the front (three panels on the left) and the rear (three panels on the right) circular blockers have on focused and ghost rays. The simulation was run for a source at infinity varying its position from on-axis to 28 arcmin off-axis (the horizontal axis of every plot). The simulated optics consisted of a single shell with physical parameters (radii and focal length) of the innermost mirror from a standard 10-shell optic. A standard detector size was used for these simulations (1cm side). We utilized source off-axis positions ranging along a direction defined by the diagonal of such detector. Plots at the top row show focused (green), and ghost rays in blue (red) coming from the paraboloid (hyperboloid) segment. All plots are normalized to the focused flux of an on-axis source. Plots on the second (third) row show singly reflected rays fluxes coming from the paraboloid (hyperboloid) segment. 
The line style for every curve corresponds to a particular set of blocker radii, as indicated at the legends. The gray dashed vertical lines indicate the minimum (left) and maximum (right) distance from the optical axis to the edge of the squared detector.}
\end{figure}

Figure \ref{fig:blockers} shows the impact of different front and rear blocker sizes on a 10-mirror optics module, as a function of the off-axis source location. The y-axis in all plots shows the intensity of rays within the detector, normalized to the maximum flux of doubly reflected rays, i.e. when the sources is on-axis. The top panels show the effect on the focused rays (green);  the panels on the left present the impact of front blocker sizes on focused rays, singly reflected rays from paraboloid segments (blue curves on the top and middle panels), and singly reflected rays from hyperboloid segments (red curves on the top and bottom panels). For all three panels on the left, front blocker radii range among 2.875~cm, 3.0967~cm, 3.1334~cm and 3.1700~cm, while there is no rear blocker. The panel at the top-left displays focused (green) and singly reflected (blue and red) rays together to show the relative flux of ghost rays when compared to focused rays. It is observed that ghost ray fluxes never surpass 35\% of the on-axis focused flux. 

According to the left panels of figure \ref{fig:blockers}, it is clear that the larger the front blocker, the less ghost rays impinge into the detector. However, the larger the front blocker, the larger is the reduction effect on the focused rays. An optimized size must balance reducing singly-reflected rays while having the smallest effect on focused rays. In this case, the best size was found to be {\bf 3.0967} cm, and {\bf 3.7494} cm radius for a 10-mirror and a 7-mirror optics module, respectively. 

The three panels on the right of figure \ref{fig:blockers} assess the effect of the rear blocker on the ghost ray background for the innermost mirror of a 10-mirror optics module. In this case, the front blocker was set to a constant radius of 3.0967 cm (optimized size found for the front blocker). These three panels on the right have an analogous structure to the left panels. For the rear blocker, the best trade-off between reducing ghost rays and minimizing vignetting effects is obtained for a radius of 2.62 cm for a 10-mirror optics. An analogous simulation was run for the innermost mirror of a standard 7-mirror optics (not shown here) leading to a radius of 3.14 cm radius as an optimal parameter for the rear circular blocker.

Blockers with optimal sizes were manufactured and implemented on the FOXSI-3 rocket payload. The effectiveness of this strategy was demonstrated with direct lab measurements shown in Figure \ref{fig:collimatorblockers} A, which shows the pattern on the focal plane generated by a 30 arcmin off-axis source shining on a 7-mirror module placed at the NASA Marshall SLF. From Figure \ref{fig:collimatorblockers} A it is concluded that ghost rays are highly mitigated within a detector (gray box on the figure) by simply implementing the right blocker sizes. For the third flight of the rocket, all optical modules were upgraded to have optimized blockers, which is now a baseline for FOXSI and should be for future Wolter-I optics.

The \code{foxsisim} toolkit was extensively utilized before the FOXSI-3 rocket campaign to ascertain efficient methods of mitigating ghost rays. In this section, it was shown that optimizing the front and rear blocker sizes is a simple way to minimize ghost rays on the detectors. Although the blockers substantially reduce ghost rays, this strategy does not eliminate all of the ghost ray background. Two additional strategies to further decrease ghost ray background are discussed next.

\subsection{Honeycomb collimator}
\label{subsubsec:honeycomb}

\begin{figure}[htbp]
\centering
\includegraphics[width=1.0\textwidth]{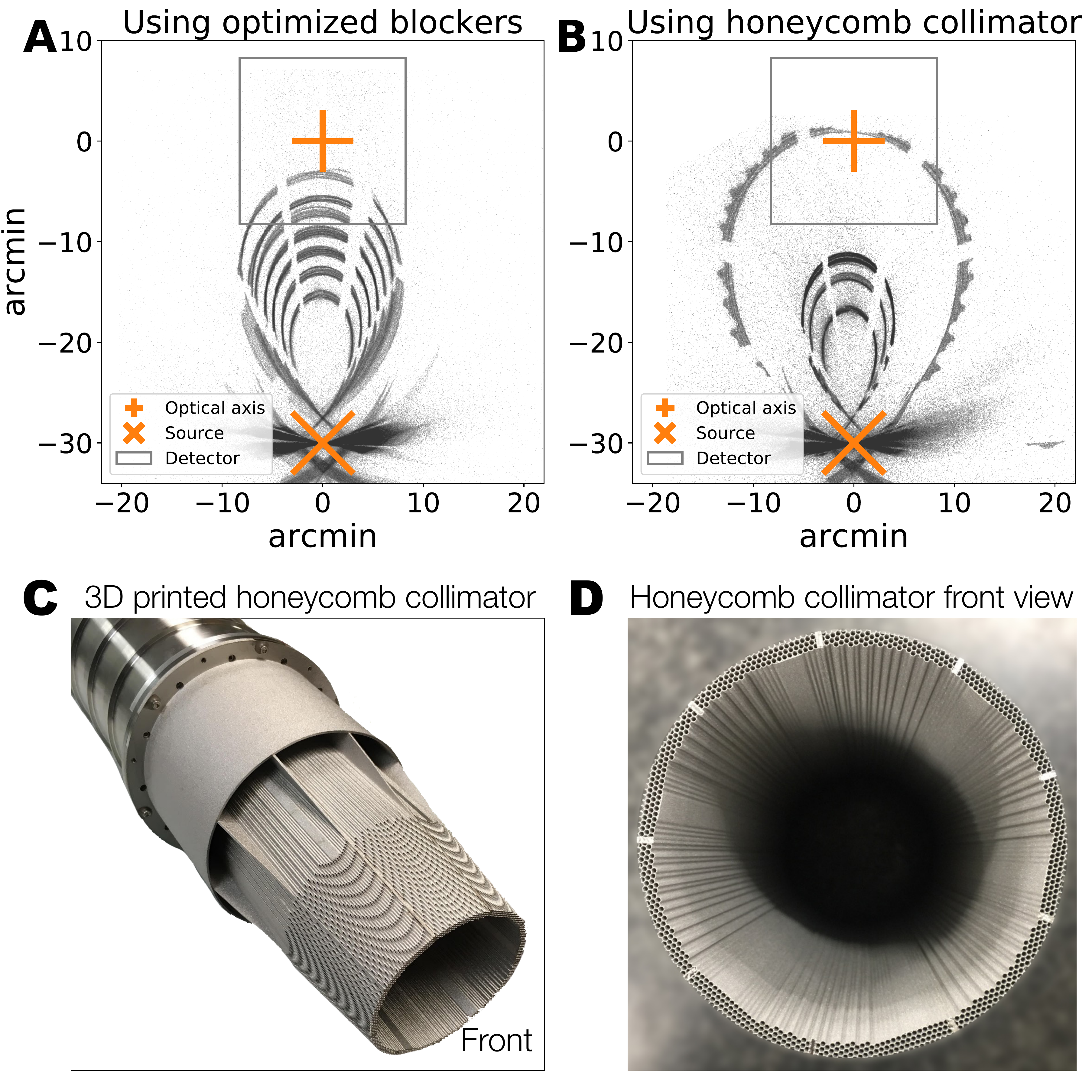}
\caption{\label{fig:collimatorblockers} Measured mitigation of ghost rays for a 30 arcmin off-axis X-ray source shining on a 7-mirror module at the NASA Marshall SLF. {\bf A} shows the ghost-rays measurement displayed over the focal plane when optimized blocker sizes are used for the optics module. {\bf B} shows how  {\bf by using blockers together with a honeycomb collimator, all ghost rays can be removed}. The thin outer-most ring on the ghost ray pattern is due to X-rays that leak through a narrow gap between the front blocker and the collimator structure. Due to mechanical constraints, we had to use a slightly smaller blocker than the one used on panel A, so that it could be physically attached to the collimator.  This gap can easily be reduced to zero to remove all ghost rays from the field of view for future missions. Panels A and B display an orange cross and {\it X} mark representing the optical axis and source location, respectively. The gray box shows a detector. {\bf C} presents a picture of the actual honeycomb collimator attached to the entrance of one 7-mirror optics module. {\bf D} shows a head-on view of the collimator and shows the honeycomb structure designed to collimate rays in front of the four innermost mirrors. Every small hexagonal hole has a 1 mm diameter and a wall thickness of 0.12 mm. The honeycomb collimator's length is 20.05 cm, which translates to an aspect ratio of up to 200.}
\end{figure}

Another method to reduce singly-reflected rays is to collimate rays before they arrive at the entrance aperture. Figures \ref{fig:GRemergence} and  \ref{fig:blockers} show that, for a 10-mirror optics, ghost rays are produced by off-axis sources beyond $\sim 13$ arcmin. For a 7-mirror optic only rays beyond $\sim 18$ arcmin make it onto the detector due to the larger graze angles. In order to remove these rays a collimator was designed and tested. A photograph of this collimator is shown in Figure \ref{fig:collimatorblockers} panel C. This honeycomb collimator was 3D printed and consists of a series of parallel, cylindrical, and uniform holes or channels through which X-rays can pass.  The longer and thinner the channels, the smaller the angle of acceptance for X-rays and therefore the greater the collimation. The thickness of the walls of the channels must be large enough stop (absorb) the pertinent X-ray energies (up to $\sim$20 keV in this case). The thicker the channel walls the larger the reduction of aperture. Since most of the detector-infringing singly- reflected rays come from the inner mirrors, the structure only collimates the four innermost mirrors of a 7-shell optical module, which helped minimize weight. For the large radii mirror shells, less collimation is necessary leading to a reduction in the required channel height at the circumference. The final version of the collimator has a maximum length of 19.5~cm, and 1~mm diameter channels with 120-micron-thick walls between the channels in a hexagonal structure. Further details on the manufacture and dimensions of this collimator can be found in section 3.2 of \citet{BuitragoCasas:2017dy}. 

The effectiveness of the collimator was tested at the MSFC SLF facility. The results can be seen in Figure \ref{fig:collimatorblockers} for a collimator optimized for a 7-mirror optic. Panels A and B compare the ghost-rays with and without the collimator for a source 30 arcmin off-axis showing a clear reduction in ghost rays. It can be seen that the collimator removes many ghost rays from the detector. The pay-off is a reduction of focused rays which depends on the open area of the collimator cross-section. The thin outermost ring observed in Figure \ref{fig:collimatorblockers} B is due to a narrow gap between the innermost part of the honeycomb structure of the collimator and the front blocker. This gap can easily be removed by adjusting the size of the blockers in future design iteration leading to an image with no ghost rays.

An important limitation of the honeycomb collimator is that this strategy does not block ghost rays near the source for long focal lengts missions (e.g. focal lengths of 10 m or longer).

% Measured Effects - Blockers & Collimator

\subsection{Wedge Absorber}
\label{subsubsec:WedgeAbsorber}

% Wedge Absorber
\begin{figure}[htbp]
\centering
\includegraphics[width=1.0\textwidth]{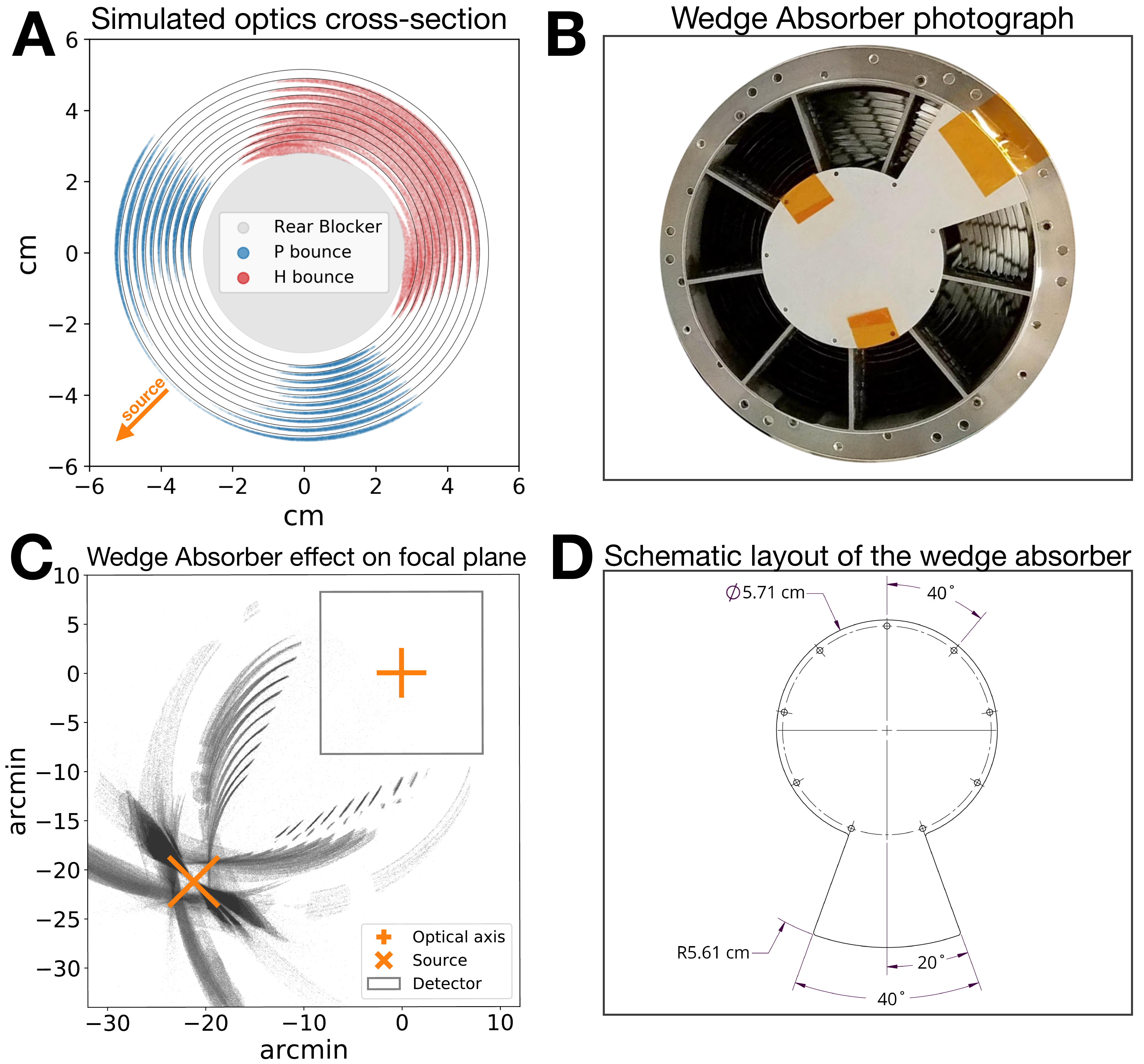}
\caption{\label{fig:wedge} A wedge absorber is a successful strategy to clear detectors of ghost ray background when a compact and intense off-axis X-ray source is present. (A) Cross-section of an optics module showing the spatial distribution of simulated singly reflected rays differentiated by color. The paraboloid (blue) and hyperboloid (red) singly reflected rays come from different regions of the optic. (B) Photograph of an aluminium 1.5 mm thick wedge absorber tightly placed at the entrance of a 10-mirror module. That optics + wedge absorber was tested at the NASA Marshall SLF. (C) Measurement of effect that the wedge absorber has on the focal plane pattern when a 30 arcmin off-axis X-ray source illuminates the module at the NASA Marshall SLF. The orange cross and {\it X} mark represent the optical axis and the location of the source respectively. The gray box represents a standard detector. As observed in panel C, ghost rays impinging a detector are negligible when implementing the wedge absorber strategy. In panel D, we display a schematic layout for the wedge absorber, made out of a 1.5 mm thick aluminum plate. The wedge disk's center needs to be placed at the optics entrance, in line with the optical axis. The wedge must be clocked according to the X-ray source's location. This wedge blocks all of the rays singly-reflected from the hyperbolic section.}
\end{figure}

The distribution of singly-reflected rays on the aperture plane suggests another method to reduce ghost rays. From Figure \ref{fig:wedge} Panel A, it is observed that singly reflected rays from the paraboloid segments (blue) and from the hyperboloid segments (red) come from different and confined angular areas of the mirrors. Since it is known that the primary source of ghost-rays are reflections from the hyperboloid segments these can be blocked at the aperture with a wedge-shaped absorber, a photograph of which can be seen in Panel B in front of a 10-mirror module. This method was tested at the MSFC SLF with a 30 arcmin off-axis source. The pattern on the focal plane is observed to be completely clear of ghost rays. The downsides of this method are that the focused rays are reduced by the wedge and that the angle of the absorber must be opposite to the polar angle of the source. This means, that it is required to know of the ghost ray source location in advance and to have control of the wedge angle.

\subsection{Comparing ghost ray mitigation strategies}
\label{sec:discussion&conclusions}

A number of strategies are discussed and compared in this section. The first strategy, blockers, was found to effectively reduce ghost rays though they must be optimized. Using the simulation, it is possible to find an optimal size for the blockers that reduces a large amount of ghost rays with minimal impact on focused rays. {\bf The use of optimized blockers should be implemented as a baseline for any Wolter-I telescope design}.

The effectiveness of the other two strategies considered in this paper to mitigate ghost rays, the wedge absorber and honeycomb collimator, are compared in Table \ref{tab:trade-study}. This trade-off study contains parameters assessing effects on the optical performance of the instrument like the efficacy at reducing ghost-rays, and degradation of the telescope effective area. We also compare the implication of each strategy on their fabrication and coupling with other parts of a potential mission. The last part of the table evaluates the possible heritage use of these two strategies for future missions.

One important limitation of the wedge absorber is that the angle of the wedge must be set in accordance with the polar angle of an off-axis source. In the right configuration, it may be possible to block ghost-rays from many sources but generally a wedge absorber can only deal with a single source. The orientation of the wedge could be tuned by either rotating the entire observatory or by adding a mechanism to rotate the wedge. A mechanism could also potentially control multiple wedges that could block multiple sources and stow under each other to return to a single wedge configuration. 

There have been several X-ray telescopes with Wolter-I mirrors that have had to deal with ghost rays. All were designed to observe astronomical objects, like Chandra, XMM Newton, and NuSTAR. The latter have been used to observe the Sun where the ghost ray situation is much more intense. Chandra and XMM Newton implemented methods to minimize, but not completely block, ghost rays. NuSTAR had some launch mass constraints that led to abandoning a collimator designed to mitigate stray light and ghost rays optimally. 

Chandra suppressed ghost rays within 14 arcmins of the optical axis by placing monolithic baffles at the very front of the telescopes and the optics' central aperture plate's aft surface. A description of the baffles is given in \citet{cusumano2007simbol}, and \citet{Chandra01}. XMM Newton also implemented X-ray baffles, but different from Chandra's. They were constructed as a series of sieve plates made out of rigid circular strips. These sieves were aligned and mounted to the front face of each telescope module. The axial space available allowed two of such sieves to be incorporated instead of monolithic baffles, blocking $\sim 80\%$ of ghost rays. The sieves designed for XMM Newton block singly reflected rays from just outside the nominal optics FOV (cone angle of 15 arcmins). A detailed description of this method is discussed in \citet{XMM-Newton}. NuSTAR is affected by stray light and ghost rays. \citet{madsen2017observational} discuss how NuSTAR ghost rays and stray light could have been prevented and should be in future observatories. For NuSTAR, some ghost ray mitigation was achieved using a compact shell spacing. \citet{madsen2017observational} indicate that to reduce further ghost rays, baffling of some sort is required.

In general, for Wolter-I telescopes with enough space between mirrors, ghost rays can be controlled by baffling within the optics, as fully described by \citet{Chandra01}. But, highly nested telescopes are challenging to baffle via this method due to the limited space between consecutive mirrors, which is the case for FOXSI. Using monolithic cylindrical baffles at the entrance of every mirror is another method to constrain ghost rays useful only for highly packed Wolter-I mirrors designed for a rather large grazing incidence angle. FOXSI's angles range from 0.23 to 0.37 degrees, with mirror spacing of 0.1-0.2 cm, this translates monolithic cylinders over 50 cm long to block ghost rays entirely. Such long cylinders would have been very challenging to manufacture and attach to each shell and would have doubled the length of the optics. For FOXSI, we also discarded sieves' use since they mitigate but not entirely block ghost rays. We had to develop new ghost ray blocking strategies, leading to the 3D honeycomb collimator and the wedge absorber presented in this paper and summarized in Table \ref{tab:trade-study}.

\begin{table}
\begin{center}
\resizebox{0.99\textwidth}{!}{%
\begin{tabular}{ | p{5cm} | p{5cm} | p{5cm} |}
\hline
{\bf Parameter} & {\bf Wedge absorber} & {\bf 3D printed honeycomb collimator} \\ \hline

Efficacy at reducing singly reflected rays & Depending on the spatial configuration of the source(s) the efficacy ranges from $\sim 0\%$ up to $\sim100\%$ & Nominally close to $100\%$ for the graze angles of the original design. \\ \hline

%Degradation of the on-axis effective area & $\sim12.5$\% & $\sim50$\%. \\ \hline

Volume and weight & Minimal & Length and mass scale inversely with threshold off-axis angle blocked.  FOXSI design (18 arcmin threshold) is 19.5 cm long, has an internal radius of 4.43 cm, an external radius of 5.3 cm, and a mass of 0.8 kg. \\ \hline

Fabrication constraints & Mechanically easy to implement, with a material dense enough to stop X-rays in the working energy range of {\it FOXSI} & 3D printing allows for easy customization of the design to minimize mass, but technology is currently limited to channels no smaller than 0.5 - 1 mm. \\ \hline

Calibration difficulties & Need to characterize the effect that the wedge absorber has on the point spread function. & It needs to be guaranteed that the optics module and collimator optical axes are co-aligned. System mis-alignments lead to a vast degradation of the telescope effective area and ghost rays.\\ \hline
% Special considerations need to be account when alignment. 
% A complete X-ray - suggests that the other calibrations are "not complete". FIX THAT.

Implications on other parts of the payload & Precise non-trivial roll control during the rocket flight. These maneuvers require careful telemetry command control on the payload roll angle. & Some implications on the location of solar pointing sensors to avoid interference. Extra difficulties at aligning the payload due to the lack of ghost rays used in such process.\\ \hline
%Say more general: The total length of the payload have to change -> changes of locations of other components.

Mechanical implications & It should be well attached to the FOXSI spider fixture to keep in place during the flight. & Minor implications mostly due to the co-alignment of the collimator mounting with a FOXSI telescope. \\ \hline

Heritage to satellite missions & Applicable with substantial modifications to allow angular mobility of the blocker. Includes extra complexity on electronics, power, and controller logic. A protocol to decide placement of the wedge need to be set. & If the honeycomb structure is reduced in size it may become an option for a spacecraft, but it can not block small off-axis anlges.\\ \hline
\end{tabular}}
\caption{Comparison of the two strategies considered to reduce singly reflected rays for the FOXSI rocket experiment.}
\label{tab:trade-study}
\end{center}
\end{table}

\section{Summary and Conclusions}
\label{sec:conclusions}

All strategies to mitigate ghost rays presented here were established by a comprehensive use of \code{foxsisim}. The mission of analysing ghost rays, and reducing them, was an excellent example to show the versatility and power of our ray tracing simulation toolbox. We present \code{foxsisim} as an open-access set of tools to study any X-ray Wolter-I optics. We encourage the broad scientific community to use the numerical tools presented here. They can be useful in applications for astronomy, medicine, material sciences, etc. Future scientific results of the quiet-Sun observations obtained during past flights of the FOXSI rocket will include the use of \code{foxsisim} to assess the overall instrument background.

We found that adding blockers optimized to maximize their ghost-ray mitigation should be a baseline for any Wolter-I telescope design. In addition, we assessed two other strategies to reduce ghost-rays. One based on a honeycomb collimator and one on a wedge absorber. The use of either of these two strategies on future spacecrafts imply some payoff that need to be assessed according to the particular design and constrains of the instrument. For example, a honeycomb collimator adds a substantial amount of mass but requires minimum, if any, maneuvering. Opposite to that, the wedge absorber is light, but requires telemetry and control of the spacecraft to locate the wedge at the right position that optimizes ghost ray mitigation. The two strategies have a direct impact on the effective area of the optics.

\acknowledgments
The first author of this paper is funded by the NASA FINESST grant 80NSSC19K1438. The FOXSI-3 sounding rocket is funded by NASA LCAS grant NNX16AL60G. UMN team members were supported by the NSF grant AGS-1429512. Japan team members were supported by JSPS KAKENHI Grant Numbers JP15H03647, JP18H03724. The FOXSI team is grateful to the NSROC teams at WSMR and Wallops for the excellent operation of their systems. Furthermore, the authors would like to acknowledge the contributions of each member of the FOXSI experiment team to the project, particularly our team members at MSFC for the fabrication of the focusing optics. Thank-you to Amy Winebarger and Patrick Champey, and Wayne Baumgartner at the MSFC for their grand disposition when we were taking data at the Stray Light Facility. We also thank Robert Taylor for initiating the writing of the \code{foxsisim} ray tracing simulation during a Summer internship at the NASA Goddard Space Flight Center. The authors thank to Mr. Hayato Inuzuka and Mr. Futoshi Yoshimura at Toray Precision Co., Ltd. for their great effort on the development of 3D printed honeycomb collimator. The first author thanks Dr. Juan Carlos Mart\'inez Oliveros for fruitful discussions during the writting of this paper.
\bibliography{references}

\begin{thebibliography}{26}
\providecommand{\natexlab}[1]{#1}
\providecommand{\url}[1]{\texttt{#1}}
\expandafter\ifx\csname urlstyle\endcsname\relax
  \providecommand{\doi}[1]{doi: #1}\else
  \providecommand{\doi}{doi: \begingroup \urlstyle{rm}\Url}\fi

\bibitem[Wolter(1950)]{wolter1950frage}
Hans Wolter.
\newblock Zur frage des lichtweges bei totalreflexion.
\newblock \emph{Zeitschrift f{\"u}r Naturforschung A}, 5\penalty0 (5):\penalty0
  276--283, 1950.

\bibitem[Cash(2002)]{cash2002medical}
Webster~C Cash.
\newblock Medical uses of focused and imaged x-rays, March~19 2002.
\newblock US Patent 6,359,963.

\bibitem[Ogasaka et~al.(2008)Ogasaka, Tamura, Shibata, Furuzawa, Miyazawa,
  Shimoda, Fukaya, Iwahara, Nakamura, Naitou,
  et~al.]{ogasaka2008characterization}
Yasushi Ogasaka, Keisuke Tamura, Ryo Shibata, Akihiro Furuzawa, Takuya
  Miyazawa, Kenta Shimoda, Yoshihiro Fukaya, Tomonaga Iwahara, Tomokazu
  Nakamura, Masataka Naitou, et~al.
\newblock Characterization of a hard x-ray telescope at synchrotron facility
  spring-8.
\newblock \emph{Japanese Journal of Applied Physics}, 47\penalty0
  (7R):\penalty0 5743, 2008.

\bibitem[Mildner and Gubarev(2011)]{mildner2011wolter}
DFR Mildner and MV~Gubarev.
\newblock Wolter optics for neutron focusing.
\newblock \emph{Nuclear Instruments and Methods in Physics Research Section A:
  Accelerators, Spectrometers, Detectors and Associated Equipment},
  634\penalty0 (1):\penalty0 S7--S11, 2011.

\bibitem[Ferreira et~al.(2013)Ferreira, Christensen, Pivovaroff, Brejnholt,
  Fernandez-Perea, Westergaard, Jakobsen, Descalle, Soufli, and
  Vogel]{ferreira2013hard}
Desiree Della~Monica Ferreira, Finn~E Christensen, Michael~J Pivovaroff,
  Nicolai Brejnholt, Monica Fernandez-Perea, Niels J{\o}rgen~S Westergaard,
  Anders~C Jakobsen, Marie-Anne Descalle, Regina Soufli, and Julia~K Vogel.
\newblock Hard x-ray/soft gamma-ray telescope designs for future astrophysics
  missions.
\newblock In \emph{Optics for EUV, X-Ray, and Gamma-Ray Astronomy VI}, volume
  8861, page 886116. International Society for Optics and Photonics, 2013.

\bibitem[Werner(1977)]{werner1977imaging}
W~Werner.
\newblock Imaging properties of wolter i type x-ray telescopes.
\newblock \emph{Applied optics}, 16\penalty0 (3):\penalty0 764--773, 1977.

\bibitem[Madsen et~al.(2017)Madsen, Christensen, Craig, Forster, Grefenstette,
  Harrison, Miyasaka, and Rana]{madsen2017observational}
Kristin~K Madsen, Finn~E Christensen, William~W Craig, Karl~W Forster, Brian~W
  Grefenstette, Fiona~A Harrison, Hiromasa Miyasaka, and Vikram Rana.
\newblock Observational artifacts of nuclear spectroscopic telescope array:
  ghost rays and stray light.
\newblock \emph{Journal of Astronomical Telescopes, Instruments, and Systems},
  3\penalty0 (4):\penalty0 044003, 2017.

\bibitem[Cusumano et~al.(2007)Cusumano, Artale, Mineo, Teresi, Pareschi, and
  Cotroneo]{cusumano2007simbol}
Giancarlo Cusumano, MA~Artale, T~Mineo, V~Teresi, Giovanni Pareschi, and
  V~Cotroneo.
\newblock Simbol-x: x-ray baffle for stray-light reduction.
\newblock In \emph{Optics for EUV, X-Ray, and Gamma-Ray Astronomy III}, volume
  6688, page 66880C. International Society for Optics and Photonics, 2007.

\bibitem[Spiga(2016)]{spiga2016analytical}
Daniele Spiga.
\newblock Analytical computation of stray light in nested mirror modules for
  x-ray telescopes.
\newblock In \emph{Optics for EUV, X-Ray, and Gamma-Ray Astronomy VII}, volume
  9603, page 96030H. International Society for Optics and Photonics, 2016.

\bibitem[Christe et~al.(2019)Christe, Buitrago-Casas, and
  Taylor]{steven_d_christe_2019_3445460}
Steven~D. Christe, Milo Buitrago-Casas, and Robert Taylor.
\newblock foxsi/foxsi-optics-sim, September 2019.
\newblock URL \url{https://doi.org/10.5281/zenodo.3445460}.

\bibitem[{Krucker} et~al.(2009){Krucker}, {Christe}, {Glesener}, {McBride},
  {Turin}, {Glaser}, {Saint-Hilaire}, {Delory}, {Lin}, {Gubarev}, {Ramsey},
  {Terada}, {Ishikawa}, {Kokubun}, {Saito}, {Takahashi}, {Watanabe},
  {Nakazawa}, {Tajima}, {Masuda}, {Minoshima}, and
  {Shomojo}]{krucker2009focusing}
Sam {Krucker}, Steven {Christe}, Lindsay {Glesener}, Steve {McBride}, Paul
  {Turin}, David {Glaser}, Pascal {Saint-Hilaire}, Gregory {Delory}, R.~P.
  {Lin}, Mikhail {Gubarev}, Brian {Ramsey}, Yukikatsu {Terada}, Shin-Nosuke
  {Ishikawa}, Motohide {Kokubun}, Shinya {Saito}, Tadayuki {Takahashi}, Shin
  {Watanabe}, Kazuhiro {Nakazawa}, Hiroyasu {Tajima}, Satoshi {Masuda}, Takashi
  {Minoshima}, and Masumi {Shomojo}.
\newblock \emph{{The Focusing Optics X-ray Solar Imager (FOXSI)}}, volume 7437
  of \emph{Society of Photo-Optical Instrumentation Engineers (SPIE) Conference
  Series}, page 743705.
\newblock SPIE, 2009.
\newblock \doi{10.1117/12.827950}.

\bibitem[Krucker et~al.(2013)Krucker, Christe, Glesener, Ishikawa, Ramsey,
  Gubarev, Saito, Takahashi, Watanabe, Tajima, et~al.]{krucker2013focusing}
S{\"a}m Krucker, Steven Christe, Lindsay Glesener, Shinnosuke Ishikawa, Brian
  Ramsey, Mikhail Gubarev, Shinya Saito, Tadayuki Takahashi, Shin Watanabe,
  Hiroyasu Tajima, et~al.
\newblock The focusing optics x-ray solar imager (foxsi): instrument and first
  flight.
\newblock In \emph{Solar Physics and Space Weather Instrumentation V}, volume
  8862, page 88620R. International Society for Optics and Photonics, 2013.

\bibitem[Christe et~al.(2016)Christe, Glesener, Buitrago-Casas, Ishikawa,
  Ramsey, Gubarev, Kilaru, Kolodziejczak, Watanabe, Takahashi, Tajima, Turin,
  Shourt, Foster, and Krucker]{christe2016JAI}
Steven~D Christe, Lindsay~Erin Glesener, Juan~Camilo Buitrago-Casas,
  Shin-Nosuke Ishikawa, Brian~D Ramsey, Mikhail~V Gubarev, Kiranmayee Kilaru,
  Jeffery~J Kolodziejczak, Shin Watanabe, Tadayuki Takahashi, Hiroyasu Tajima,
  Paul Turin, Van Shourt, Natalie Foster, and Sam Krucker.
\newblock {FOXSI-2: Upgrades of the Focusing Optics X-ray Solar Imager for its
  Second Flight}.
\newblock \emph{Journal of Astronomical Instrumentation}, 5\penalty0
  (1):\penalty0 1640005--1640625, March 2016.

\bibitem[{Ramsey} et~al.(2002){Ramsey}, {Alexander}, {Apple}, {Benson},
  {Dietz}, {Elsner}, {Engelhaupt}, {Ghosh}, {Kolodziejczak}, {O'Dell},
  {Speegle}, {Swartz}, and {Weisskopf}]{Ramsey2002}
Brian~D. {Ramsey}, Cheryl~D. {Alexander}, Jeff~A. {Apple}, Carl~M. {Benson},
  Kurtis~L. {Dietz}, Ronald~F. {Elsner}, Darell~E. {Engelhaupt}, Kajal~K.
  {Ghosh}, Jeffery~J. {Kolodziejczak}, Stephen~L. {O'Dell}, Chet~O. {Speegle},
  Douglas~A. {Swartz}, and Martin~C. {Weisskopf}.
\newblock {First Images from HERO, a Hard X-Ray Focusing Telescope}.
\newblock \emph{Astrophysical Journal}, 568\penalty0 (1):\penalty0 432--435,
  Mar 2002.
\newblock \doi{10.1086/338801}.

\bibitem[{Ramsey}(2005)]{Ramsey2005}
Brian~D. {Ramsey}.
\newblock {Replicated Nickel Optics for the Hard-X-Ray Region}.
\newblock \emph{Experimental Astronomy}, 20\penalty0 (1-3):\penalty0 85--92,
  Dec 2005.
\newblock \doi{10.1007/s10686-006-9033-6}.

\bibitem[Krucker et~al.(2014)Krucker, Christe, Glesener, Ishikawa, Ramsey,
  Takahashi, Watanabe, Saito, Gubarev, Kilaru, et~al.]{krucker2014first}
S{\"a}m Krucker, Steven Christe, Lindsay Glesener, Shin-nosuke Ishikawa, Brian
  Ramsey, Tadayuki Takahashi, Shin Watanabe, Shinya Saito, Mikhail Gubarev,
  Kiranmayee Kilaru, et~al.
\newblock First images from the focusing optics x-ray solar imager.
\newblock \emph{The Astrophysical Journal Letters}, 793\penalty0 (2):\penalty0
  L32, 2014.

\bibitem[Glesener et~al.(2016)Glesener, Krucker, Christe, Ishikawa,
  Buitrago-Casas, Ramsey, Gubarev, Takahashi, Watanabe, Takeda, Courtade,
  Turin, McBride, Shourt, Hoberman, Foster, and Vievering]{Glesener:2016eq}
Lindsay~Erin Glesener, Sam Krucker, Steven~D Christe, Shin-Nosuke Ishikawa,
  Juan~Camilo Buitrago-Casas, Brian~D Ramsey, Mikhail~V Gubarev, Tadayuki
  Takahashi, Shin Watanabe, Shinichiro Takeda, Sasha Courtade, Paul Turin,
  Stephen McBride, Van Shourt, Jane Hoberman, Natalie Foster, and Juliana
  Vievering.
\newblock {The FOXSI solar sounding rocket campaigns}.
\newblock In Jan-Willem~A den Herder, Tadayuki Takahashi, and Marshall Bautz,
  editors, \emph{SPIE Astronomical Telescopes + Instrumentation}, page 99050E.
  SPIE, July 2016.

\bibitem[Musset et~al.(2019)Musset, Buitrago-Casas, Glesener, Bongiorno,
  Courtade, Athiray, Vievering, Ishikawa, Narukage, Furukawa, Ryan, Dalton,
  Turin, Davis, Takahashi, Watanabe, Mitsuishi, Hagino, Kawate, Turin, Christe,
  Ramsey, and Krucker]{Musset:2019gj}
Sophie Musset, Juan~Camilo Buitrago-Casas, Lindsay~Erin Glesener, Stephen
  Bongiorno, Sasha Courtade, P~Subramania Athiray, Juliana Vievering,
  Shin-Nosuke Ishikawa, Noriyuki Narukage, Kento Furukawa, Daniel~F Ryan, Greg
  Dalton, Zoe Turin, Lance Davis, Tadayuki Takahashi, Shin Watanabe, Ikuyuki
  Mitsuishi, Kouichi Hagino, Tomoko Kawate, Paul Turin, Steven~D Christe,
  Brian~D Ramsey, and Sam Krucker.
\newblock {Ghost-ray reduction and early results from the third FOXSI sounding
  rocket flight}.
\newblock In Oswald~H Siegmund, editor, \emph{UV, X-Ray, and Gamma-Ray Space
  Instrumentation for Astronomy XXI}, page~37. SPIE, September 2019.

\bibitem[Athiray et~al.(2017)Athiray, Buitrago-Casas, Bergstedt, Vievering,
  Musset, Ishikawa, Glesener, Takahashi, Watanabe, Courtade,
  et~al.]{athiray2017calibration}
PS~Athiray, Juan~Camilo Buitrago-Casas, Kendra Bergstedt, Juliana Vievering,
  Sophie Musset, Shin-nosuke Ishikawa, Lindsay Glesener, Tadayuki Takahashi,
  Shin Watanabe, Sasha Courtade, et~al.
\newblock Calibration of the hard x-ray detectors for the foxsi solar sounding
  rocket.
\newblock In \emph{UV, X-Ray, and Gamma-Ray Space Instrumentation for Astronomy
  XX}, volume 10397, page 103970A. International Society for Optics and
  Photonics, 2017.

\bibitem[Furukawa et~al.(2019)Furukawa, Buitrago-Casas, Vievering, Hagino,
  Glesener, Athiray, Krucker, Watanabe, Takeda, Ishikawa,
  et~al.]{furukawa2019development}
Kento Furukawa, Juan~Camilo Buitrago-Casas, Juliana Vievering, Kouichi Hagino,
  Lindsay Glesener, PS~Athiray, S{\"a}m Krucker, Shin Watanabe, Shin’ichiro
  Takeda, Shin’nosuke Ishikawa, et~al.
\newblock Development of 60 $\mu$m pitch cdte double-sided strip detectors for
  the foxsi-3 sounding rocket experiment.
\newblock \emph{Nuclear Instruments and Methods in Physics Research Section A:
  Accelerators, Spectrometers, Detectors and Associated Equipment},
  924:\penalty0 321--326, 2019.

\bibitem[Buitrago-Casas et~al.(2017)Buitrago-Casas, Glesener, Christe, Ramsey,
  Elsner, Courtade, Vievering, Subramania, Krucker, Ishikawa, Narukage, Turin,
  and Musset]{BuitragoCasas:2017dy}
Juan~Camilo Buitrago-Casas, Lindsay~Erin Glesener, Steven~D Christe, Brian~D
  Ramsey, Ronald Elsner, Sasha Courtade, Juliana Vievering, Athiray Subramania,
  Sam Krucker, Shin-Nosuke Ishikawa, Noriyuki Narukage, Paul Turin, and Sophie
  Musset.
\newblock {Methods for reducing singly reflected rays on the Wolter-I focusing
  mirrors of the FOXSI rocket experiment}.
\newblock In Giovanni Pareschi and Stephen~L O'Dell, editors, \emph{Optics for
  EUV, X-Ray, and Gamma-Ray Astronomy VIII}, pages 18--16. SPIE, September
  2017.

\bibitem[Ishikawa et~al.(2017)Ishikawa, Glesener, Krucker, Christe,
  Buitrago-Casas, Narukage, and Vievering]{ishikawa2017detection}
Shin-nosuke Ishikawa, Lindsay Glesener, S{\"a}m Krucker, Steven Christe,
  Juan~Camilo Buitrago-Casas, Noriyuki Narukage, and Juliana Vievering.
\newblock Detection of nanoflare-heated plasma in the solar corona by the
  foxsi-2 sounding rocket.
\newblock \emph{Nature Astronomy}, 1\penalty0 (11):\penalty0 771--774, 2017.

\bibitem[Athiray et~al.(2020)Athiray, Vievering, Glesener, Ishikawa, Narukage,
  Buitrago-Casas, Musset, Inglis, Christe, Krucker, et~al.]{athiray2020foxsi}
PS~Athiray, Juliana Vievering, Lindsay Glesener, Shin-nosuke Ishikawa, Noriyuki
  Narukage, Juan~Camilo Buitrago-Casas, Sophie Musset, Andrew Inglis, Steven
  Christe, S{\"a}m Krucker, et~al.
\newblock Foxsi-2 solar microflares. i. multi-instrument differential emission
  measure analysis and thermal energies.
\newblock \emph{The Astrophysical Journal}, 891\penalty0 (1):\penalty0 78,
  2020.

\bibitem[Attwood and Sakdinawat(2017)]{attwood2017x}
David Attwood and Anne Sakdinawat.
\newblock \emph{X-rays and extreme ultraviolet radiation: principles and
  applications}.
\newblock Cambridge university press, 2017.

\bibitem[Gaetz et~al.(2000)Gaetz, Jerius, Edgar, Van~Speybroeck, Schwartz,
  Markevitch, Taylor, and Schulz]{Chandra01}
Terrance~J Gaetz, Diab Jerius, Richard~J Edgar, Leon~P Van~Speybroeck, Daniel~A
  Schwartz, Maxim~L Markevitch, SC~Taylor, and Norbert~S Schulz.
\newblock Orbital verification of the cxo high-resolution mirror assembly
  alignment and vignetting.
\newblock In \emph{X-Ray Optics, Instruments, and Missions III}, volume 4012,
  pages 41--52. International Society for Optics and Photonics, 2000.

\bibitem[de~Chambure et~al.(1999)de~Chambure, Laine, van Katwijk, Ruehe,
  Schink, Hoelzle, Gutierrez, Domingo, Ibarretxe, Tock, et~al.]{XMM-Newton}
Daniel de~Chambure, Robert Laine, Kees van Katwijk, Wolfgang Ruehe, Dietmar
  Schink, Edgar Hoelzle, Yolanda Gutierrez, Miquel Domingo, Inigo Ibarretxe,
  Jean~Philippe Tock, et~al.
\newblock X-ray baffle of the xmm telescope: development and results.
\newblock In \emph{Design and Engineering of Optical Systems II}, volume 3737,
  pages 396--408. International Society for Optics and Photonics, 1999.

\end{thebibliography}

\newpage
\appendix
\section{Appendix: The effect of ghost rays on potential instrument concepts with long focal lengths}
\label{appendix}

\begin{figure}[htbp]
\centering
\includegraphics[width=1.0\textwidth]{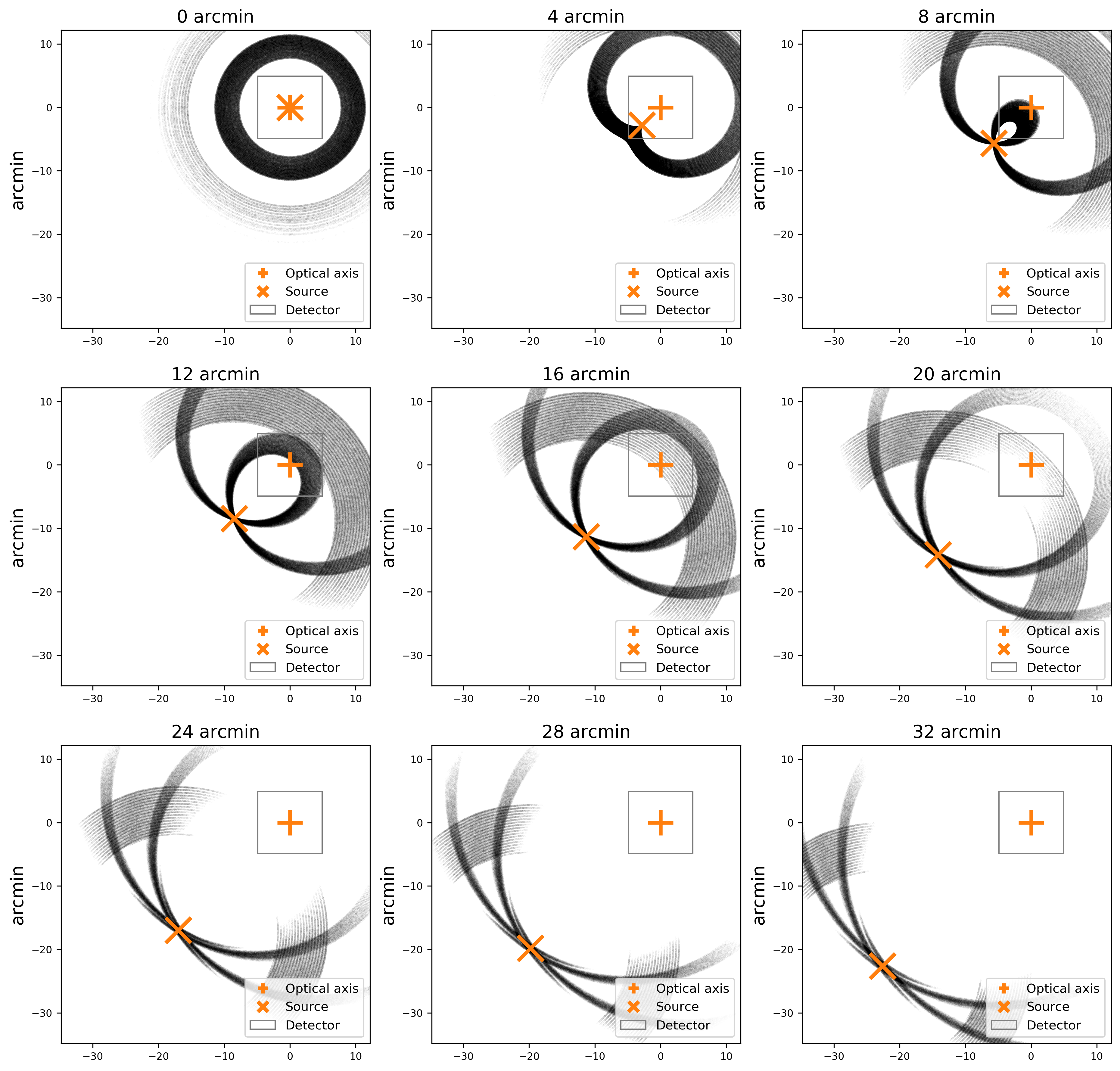}
\caption{\label{fig:GRemergence_SMEX}
Simulated ghost ray images for a 14-m focal length 18-shell telescope module that might be appropriate for a future space-based x-ray observatory whose science objectives include investigating the plasma heating and acceleration processes in solar flares. 
Ghost ray patterns are simulated for a point source at infinity with off-axis angles from 0 to 32 arcmin. The gray square shows a 4 cm$\times$4 cm field of view. The orange symbols show the optical axis and source position. 
For a source on-axis, no ghost-rays infringe the focal plane.
Single bounce rays from the parabolic mirror segment form the inner circular pattern.
Straight-through rays form the outer circular pattern.
Sources with off-axis angles greater than $\approx$20~arcmin do not contribute any ghost rays to the imaged focal plane due to the blocking effect of the closely-packed mirrors.
For comparison, the Sun is approximately 30 arcmin across.
}
\end{figure}

The analysis presented in this paper is specific to the FOXSI sounding rocket optics configuration whose focal length (2 m) is limited by the capabilities of modern sounding rockets.
In this section, we expand upon this analysis to a configuration that might be appropriate for a future space-based x-ray observatory whose science objectives include investigating the plasma heating and acceleration processes in solar flares. 
Such an observatory would need to observe both the thermal and non-thermal emission in large flares.
The transition between thermal and non-thermal emission in a large flare occurs around 30~keV for some of the biggest flares.
Above this energy, the spectrum becomes a power law.
To determine the slope of the non-thermal power law, in order to understand the acceleration and transport processes of high energy electrons, requires observations up to $\approx$50~keV.
This high energy requirement drives the need for a relatively long focal length of at least 14 meters.
To provide sufficient effective area to detect the signature of the acceleration mechanism requires a significant number of mirror shells.
The following analysis assumes a Wolter-I optics configuration with 18 mirror shells with (intersection) radii ranging from 9.3 to 6.3~cm and graze angles from 0.1 to 0.06~deg.
The spacing between the shells has been optimized to minimize ghost rays.
A detector area of 4~cm$\times$4~cm is assumed.
The simulated ghost ray pattern is shown in Figure~\ref{fig:GRemergence_SMEX}.
In this configuration, the ghost ray patterns are significantly different than those presented in Figure~\ref{fig:GRemergence} primarily due to the long focal length or, equivalently, small graze angles.
For a source on-axis, no ghost-rays infringe the focal plane.
An inner circle pattern surrounds the focal plane which come from the parabolic segment.
The outer circle pattern is caused by rays which go straight between the mirror shells.
These rays are much less intense since they are not ``focused'' by the mirror shells.
As the source moves off-axis, there is a point at which the properly-reflected source and its ghost rays are observed together and at the same location.
Further increasing the off-axis angles leads to ghost rays patterns moving across the focal plane.
Sources with off-axis angles greater than $\approx$20~arcmin do not contribute any ghost rays to the imaged focal plane due to the blocking effect of the closely-packed mirrors.
For comparison, the Sun is approximately 30 arcmin across.
The central 6 arcmin diameter of the field of view is completely free of any ghost rays.
This is sufficiently large to contain the emission of an entire eruptive event including typical observatory pointing control requirements.

\begin{figure}[htbp]
\centering % \begin{center}/\end{center} takes some additional vertical space
\includegraphics[width=1.0\textwidth]{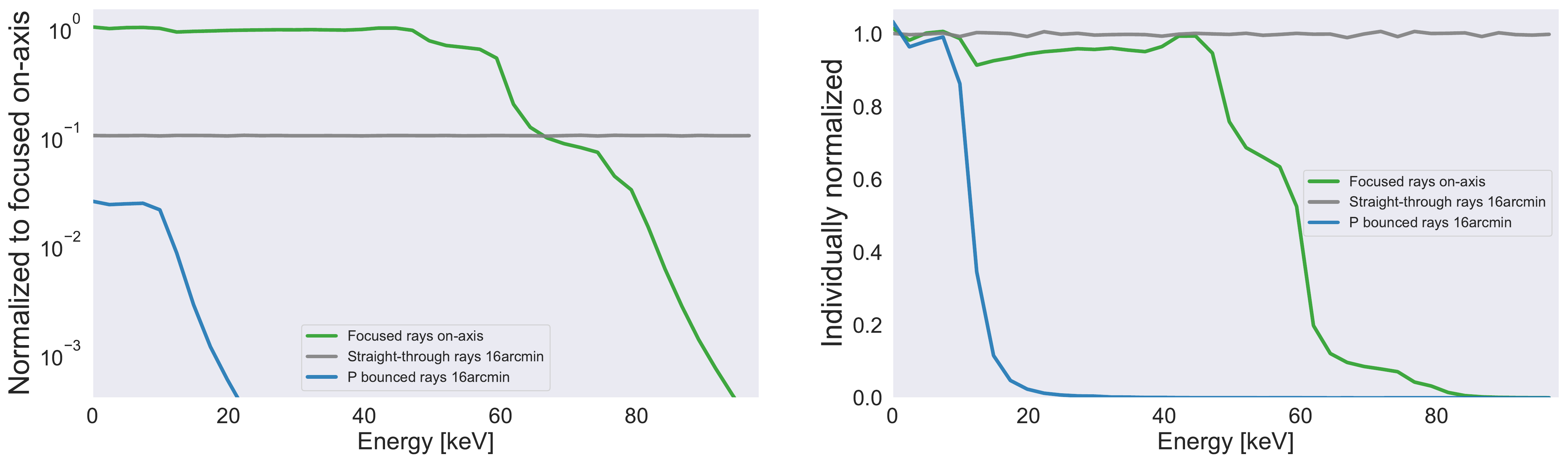}
\caption{\label{fig:GRenergy_SMEX}
A comparison of the flux of properly focused rays from an on-axis source with a flat spectrum compared to ghost rays by a source of the same brightness at an off-axis angle of 16~arcmin.
The green line the spectrum of properly focused rays.
The grey line shows the spectrum of the straight through rays which simply show the input spectrum since these rays do not interact with the mirrored surfaces.
The blue line shows that the flux of the ghost rays which are significantly attenuated by a factor of $>20$ up to 10~keV.
Above this energy, the ghost rays are increasingly attenuated compared to the on-axis focused source; at 20~keV the ghost rays are 1000 times weaker than the same source on axis.}
\end{figure}

Figure~\ref{fig:GRenergy_SMEX} shows a comparison of the flux of properly focused rays from an on-axis source with a flat spectrum compared to ghost rays by a source of the same brightness at an off-axis angle of 16~arcmin.
The green line shows that rays up to $\approx$50~keV are focused by this prescription.
The grey line shows the spectrum of the straight through rays which simply show the input spectrum since these rays do not interact with the mirrored surfaces.
The blue line shows that the flux of the ghost rays are significantly attenuated by a factor of $>20$ up to 10~keV. 
Above this energy, the ghost rays are increasingly attenuated compared to an on-axis focused source; at 20~keV the ghost rays are 1000 times weaker than the same source on axis.
This means that ghost-rays will not significantly affect observations of bright sources like solar flares or even weaker sources such as active regions as long as the sources outside of the field of view are of equal intensity.
For times when the sources outside of the field of view are bright and generate ghost rays comparable in flux to the source in the field view then more advanced deconvolution techniques can be utilized to recover faint sources.

%\subsection{Ghost rays emergence}

\end{document}